\documentclass[]{emulateapj}

\def\HI{H{\small I}\ }

\begin{document}
\title{Improving Foreground Subtraction in Statistical Observations of 21 cm Emission from the Epoch of Reionization}
\author{Miguel F. Morales\altaffilmark{1}, Judd D. Bowman\altaffilmark{2}, \& Jacqueline N. Hewitt\altaffilmark{2}}

\email{mmorales@space.mit.edu; jdbowman@mit.edu}

\altaffiltext{1}{Harvard-Smithsonian Center for Astrophysics, 60 Garden Street, MS-51 P-225, Cambridge MA 02138-1516}
\altaffiltext{2}{MIT Kavli Institute for Astrophysics and Space Research, 77 Massachusetts Ave., Cambridge, MA 02139}

\begin{abstract}
Statistical observations of the Epoch of Reionization using the 21 cm line of neutral hydrogen have the potential to revolutionize our understanding of structure formation and the first luminous objects.  However, these observations are complicated by a host of strong foreground sources. Several foreground removal techniques have been proposed in the literature, and it has been assumed that these would be used in combination to reveal the Epoch of Reionization (EOR) signal. By studying the characteristic subtraction errors of the proposed foreground removal techniques, we identify an additional subtraction stage that can further reduce the EOR foreground contamination, and study the interactions between the foreground removal algorithms.  This enables us to outline a comprehensive foreground removal strategy that incorporates all previously proposed subtraction techniques. Using this foreground removal framework and the characteristic subtraction errors,  we discuss the complementarity of different foreground removal techniques and the implications for array design and the analysis of EOR data.
\end{abstract}

\section{Introduction}

Highly redshifted 21 cm neutral hydrogen emission from the Epoch of Reionization (EOR) is a unique cosmological probe, and observations with planned low frequency radio telescopes could revolutionize our understanding of galaxy and structure formation and the emergence of the first luminous objects.  

The potential of 21 cm observations was first recognized by \citet{SunyaevZeldovich} and further developed by \citet{ScottRees,MMR97,Tozzi,Iliev}.  There are five possible experimental signatures produced by neutral hydrogen during the EOR which can be targeted with low frequency radio observations:  the global frequency step \citep{1999A&A...345..380S}, direct imaging, \HI forrest \citep {Carilli21Absorption,2004NewAR..48.1053C}, Stromgren sphere mapping, and statistical observations (see \citet{Carilli21cmSKAOverview} for overview).  Of the five experimental signatures, the statical observations developed by \citet{ZaldarriagaPow1}, \citet{MoralesEOR1}, and \citet{BharadwajVis} offer the most promise and cosmological power, and are being targeted by the Mileura Widefield Array (MWA), the LOw Frequency ARray (LOFAR), and the Chinese Meter Array (21CMA, formerly PAST).

Unlike the Cosmic Microwave Background (CMB) emission, which is inherently two dimensional (sky position), the EOR data is three dimensional  because the redshift of the observed neutral hydrogen emission maps to the line-of-sight distance. This allows us to extend the statistical techniques developed for the CMB to three dimensions, and use power spectrum statistics to study the EOR.  These statistical analysis techniques dramatically increase the sensitivity of first generation EOR observations, and allow much more detailed studies of the cosmology \citep{BowmanEOR3,FurlanettoPS1,MoralesEOR2}. The major remaining question of EOR observations is whether the foreground contamination---which is $\sim$5 orders of magnitude brighter than the neutral hydrogen radio emission---can be removed with the precision needed to reveal the underlying EOR signal.  

Several foreground subtraction techniques have been suggested in the literature.  Bright sources can be identified and removed as in CMB and galaxy clustering analyses, but the faint emission of sources below the detection threshold will still overwhelm the weak EOR signal \citep{DiMatteoForegrounds}. Additional foreground contamination can be removed by fitting a spectral model to each pixel in the sky to remove the many faint continuum sources in each line of sight \citep{BriggsForegroundSub,WangMaxForeground}.  A similar technique proposed by  \citet{ZaldarriagaPow1} moves to the visibility space (2D FT of image--frequency cube to obtain wavenumbers in sky coordinates and frequency along the third axis) and then fits smooth spectral models for each visibility \citep{2005ApJ...625..575S}. This method should be better at removing emission on larger angular scales, such as continuum emission from our own Milky Way. \citet{MoralesEOR1} introduced a subtraction technique which exploits the difference between the spherical symmetry of the EOR power spectrum and separable-axial symmetry of the foregrounds in the three dimensional Fourier space, and  is particularly well suited for removing very faint contaminants. 

Foreground removal has been envisioned as a multi-staged process in which increasingly faint contaminants are subtracted in a stepwise fashion.  By studying the errors made by proposed foreground subtraction algorithms, we identify an additional subtraction stage where the average fitting errors of the proposed algorithms are subtracted from the three dimensional power spectrum. This residual error subtraction step can significantly reduce the residual foreground contamination of the EOR signal, and differs from CMB techniques by relying on the statistics of the errors and separating the residual contamination from the power spectrum instead of the image.

Because the residual error subtraction relies on the statistical characteristics of the subtraction errors, the foreground removal steps become tightly linked and we must move from focusing on individual subtraction algorithms to the context of a complete foreground removal framework. This paper outlines a comprehensive foreground removal strategy that incorporates all previously proposed subtraction techniques and introduces the new residual error subtraction stage. Treating the foreground removal process as a complete system also allows us to study the interactions (covariance) of the subtraction algorithms and identify the types of foreground contamination each algorithm is best suited for.

Section \ref{Experiments} reviews the properties of the data produced by the first generation EOR observatories.  We then introduce the foreground removal framework in Section \ref{SubtractionStagesSec} along with several detailed examples of the subtraction errors.  Sections \ref{ConstraintsSection}  and \ref{ForegroundModels} then discuss the implications of the foreground removal framework and how it can be used to improve the design of EOR observatories and foreground removal algorithms.

\section{Experimental Data}
\label{Experiments}

While there are important differences in the data processing requirements of the MWA, LOFAR, and 21CMA---and their data analysis systems are rapidly evolving---all three experiments follow the same basic data reduction strategy. 

All three observatories are composed of thousands of simple detection elements (dual polarization dipoles for MWA and LOFAR, and single-polarization YAGIs for 21CMA).  The signals of the individual detecting elements are then combined with analog and digital systems into ``antennas'' of tens to hundreds of elements, which are then cross-correlated to produce the visibilities of radio astronomy.  These visibilities are the basic observable, and are the spatial Fourier transform of the brightness distribution on the sky at each frequency. The visibilities from each experiment---up to 4 billion per second in the case of the MWA---must then be calibrated and integrated to form one long exposure.  The final data product is a visibility cube representing a few hundred hours of observation, which can be either Fourier transformed along the angular dimensions to produce an image cube for mapping, or along the frequency axis to produce the Fourier representation for the power spectrum analysis.

Going from the raw visibilities produced by the correlators to visibility cubes representing hundreds of hours of observation is a Herculean task, and we do not wish to minimize the effort involved in this stage of the processing.  The ionospheric distortion must be corrected using radio adaptive optics, and the time variable gain, phase, and polarization of each antenna must be precisely calibrated.  Going from the raw visibilities to the long integration visibility cube tests and displays the art of experimental radio astronomy. However, in the end all three experiments will produce the same basic data product, and for our purposes we will concentrate on how to process this long integration visibility cube to remove the astrophysical foregrounds and reveal the cosmological EOR signal.

\section{Foreground Subtraction Framework}
\label{SubtractionStagesSec}
Fundamentally, all the proposed foreground subtraction techniques exploit symmetry differences between the foregrounds and the EOR signal, and are targeted at removing different types of foreground contamination.  Because the EOR signal is created by a redshifted line source, the observed frequency can be mapped to the line-of-sight distance.  This produces a cube of space where we can observe the \HI intensity as a function of position.  The EOR emission appears as bumps along both the frequency and angular directions, and since space is isotropic (rotationally invariant) leads to a spherical symmetry in the Fourier space (the 3D Fourier Transform of the image cube \citep{MoralesEOR1}).  This contrasts with most of the foreground sources which either have continuum emission which is very smooth in the frequency direction, such as synchrotron radiation, or emission line radiation which is not redshifted and thus at set frequencies, such as radio recombination lines from the Milky Way. The foreground removal techniques all use the difference between the clumpy-in-all-directions EOR signal and the foregrounds which are smooth in at least one of the dimensions. 
\begin{figure*}
\begin{center}
\plottwo{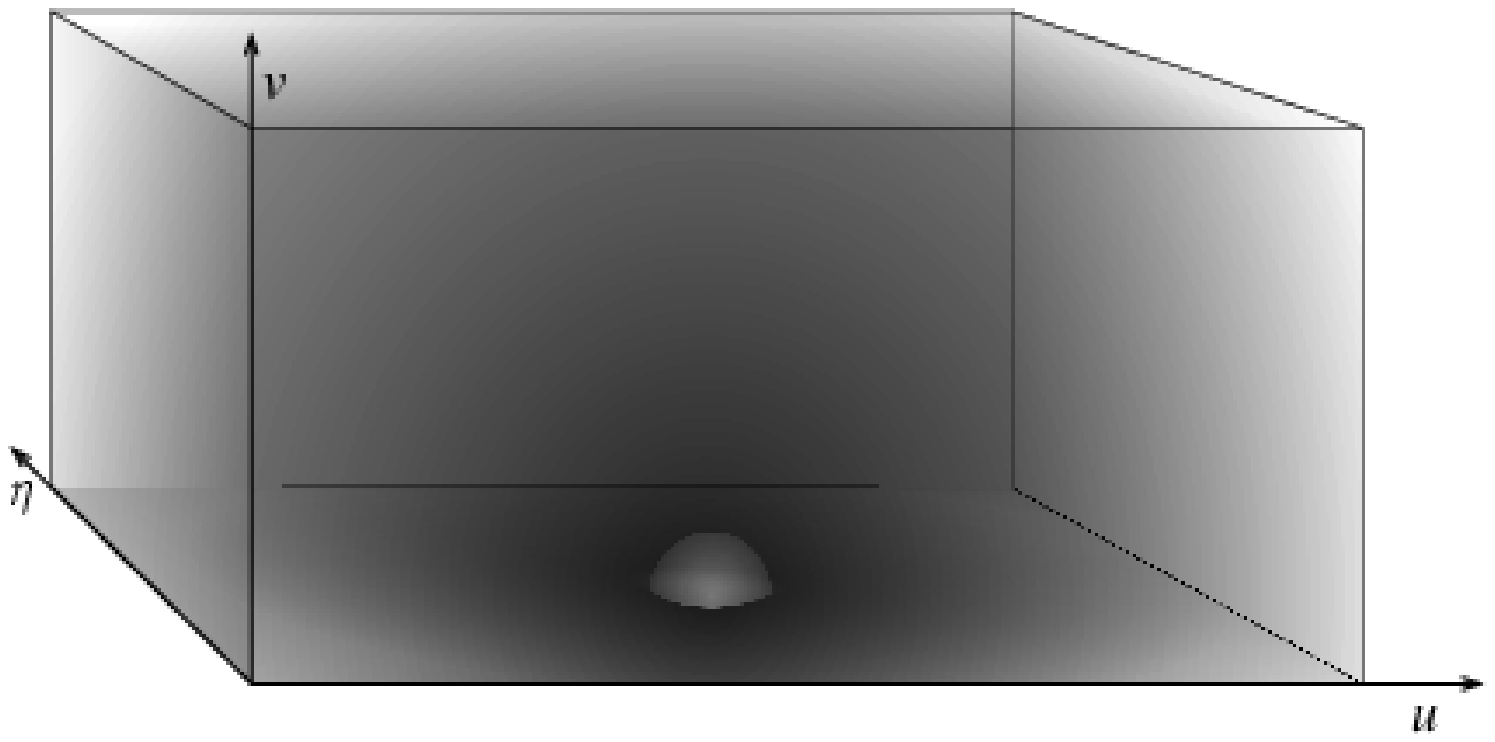}{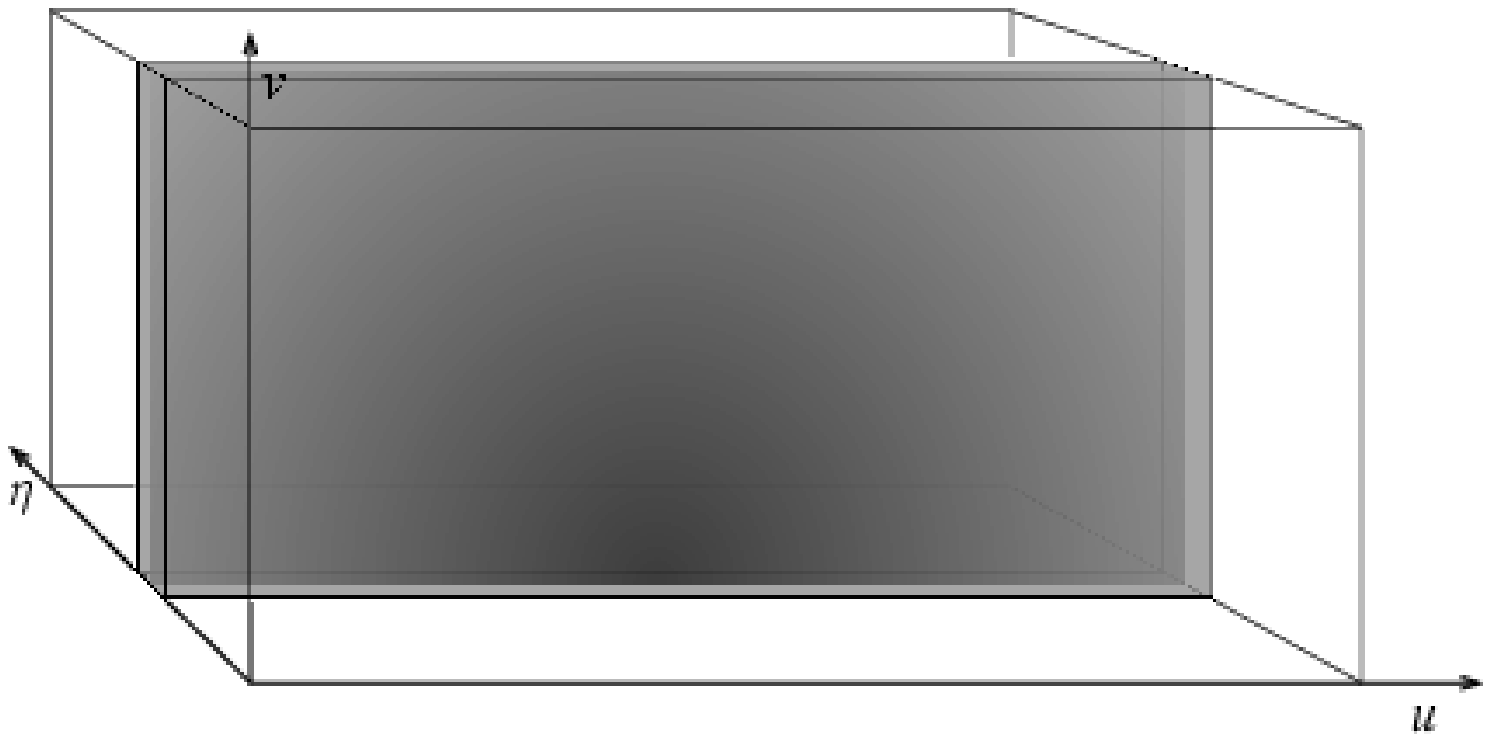}
\caption{The left panel shows the spherically symmetric power spectrum of the EOR signal (zero is in the center of the lower face), while panel (b) shows the separable-axial power spectrum template typical of the residual foregrounds.  The power spectrum shapes are known, and the amplitudes can be fit in the residual error subtraction stage to separate the residual foreground subtraction errors from the faint EOR signal.}
\label{symmetryFig}
\end{center}
\end{figure*}

The foreground subtraction framework can be divided into three stages---bright source removal, spectral fitting, and residual error subtraction---where each step subtracts increasingly faint foreground contamination.  The first two steps utilize well developed radio analysis techniques and have been previously proposed.  The residual error subtraction stage extends the symmetry ideas of \citet{MoralesEOR1} to identifying and removing the average fitting error of the first two stages.  In Sections \ref{BrightSourceSec}--\ref{RESubtraction} we step through the three subtraction stages. Section \ref{SubtractionErrorsSection} then shows several examples of how to calculate the characteristics of the subtraction errors, and Section \ref{UncertaintySec} discusses how the subtraction errors affect the EOR sensitivity.
  
One subtlety that arises is which sources should be considered foreground in an interferometric observation.  Diffuse synchrotron emission from our own galaxy is the single brightest source of radio emission at EOR frequencies.  The brightness temperature of the synchrotron emission towards the Galactic poles is several hundred degrees, and dominates the thermal noise of the telescope and receiver system. However, the diffuse Galactic synchrotron emission is so spatially and spectrally smooth it is not customarily included in discussions of foreground subtraction---the majority of this nearly DC emission is resolved out by interferometric observations.  Instead the Galactic synchrotron contribution is included in sensitivity calculations as the dominant source of system noise \citep{BowmanEOR3}. Similarly, polarized emission can become a major foreground \citep{HaverkornPolarization} by leaking into the intensity maps.  However, since the contamination of the intensity map is through errors in the polarization calibration, this is customarily considered a calibration issue and not foreground subtraction. Following these conventions here, we focus on removing the unpolarized contributions from resolved foreground sources.

\subsection{Bright Source Removal}
\label{BrightSourceSec}

In the first stage the bright contaminating sources, both astrophysical and man made, are removed.  Because the spatial and frequency response of an array is not a delta-function, emission from a bright source will spill over into neighboring pixels and frequency channels.  The goal of the first foreground removal stage is to subtract the contributions from all sources which can contaminate distant locations in the image cube. 

The worst of the radio frequency interference (RFI) will be cut out of the data prior to forming the long integration visibility cube.  What will remain is a sea of faint transmissions.  The easiest way to remove narrow-band transmissions is to identify the affected channels (elevated rms) and excise them.  Modern polyphase filters have very high dynamic range, so only a few channels will need to be removed for all but the very brightest transmitters.  This leads to slices of missing frequency information and complicates the experimental window function, but is very effective if the dynamic range of the polyphase filters is high enough.

Removing astrophysical sources is conceptually similar, but is more difficult due to the lower spatial dynamic range of most radio arrays.  Emission from a bright astrophysical source will leak into pixels far from the source position due to the imperfect point spread function of an array.  Thus we need to use the traditional radio astronomy subtraction technique of removing the sources directly from the visibilities to subtract the array sidelobes along with the central emission.  

The signal strength of the EOR is a few mK, so the astrophysical and RFI sources must be subtracted until the sidelobes are $\lesssim$mK.  This places strong constraints on the spatial and frequency dynamic range of an array, as well as the RFI environment. Unfortunately, even the faint emission of galaxies below the detection threshold will overwhelm the weak EOR signal \citep{DiMatteoForegrounds}, and we must resort to more powerful subtraction techniques to reveal the EOR signal.

\subsection{Spectral Fitting}
\label{SpectralFitSec}

At the end of the bright source foreground removal stage, all sources bright enough to corrupt distant areas of the image cube have been removed, and we are left with a cube where all of the contamination is local.  Here we can use foreground subtraction techniques which target the frequency characteristics of the foreground emission.

In every pixel of the image cube, there will be contributions from many faint radio galaxies.  The spectrum within one pixel is well approximated by a power-law, and can be fit and removed.  Since the EOR signal is bumpy, fitting smooth power law models nicely removes the foreground contribution while leaving most of the cosmological signal \citep{BriggsForegroundSub}.  There are a number of subtle effects which must be carefully monitored, such as changing pixel size, but this is an effective way of removing the contributions of the faint radio galaxy foreground.

A similar method was proposed by \citet{ZaldarriagaPow1}, where smooth spectral models are fit to individual spatial frequency pixels in the visibility space.  While there is a lot of overlap between this method and image method, the visibility foreground subtraction technique should be superior for more extended objects such as fluctuations in the Milky Way synchrotron emission.

The last type of spectral fitting is to remove radio recombination lines from our own Galaxy.  The intensity of these lines is uncertain, but since they occur at known frequencies, template spectra can be used to subtract them. Unlike the smooth power-law spectra, the structure in the recombination line spectrum has much more power on small scales (line-of-sight redshift distance). More work is needed to accurately determine the strength of this foreground and develop template spectra.

The errors made in the spectral fitting stage can be classified into two types:  \textit{model errors} due to foreground spectra which cannot be fit exactly by the model parameters, and \textit{statistical errors} due to slight misestimates of the model parameters in the presence of thermal noise.  These errors are discussed at length in Section \ref{SubtractionErrorsSection}.

\subsection{Residual Error Subtraction}
\label{RESubtraction}

While the vast majority of the foreground contamination will be removed in the first two analysis stages, residual foreground contamination will remain due to errors in the subtraction process. In the absence of foregrounds the EOR power spectrum could be measured by dividing the individual power measurements in the Fourier space into spherical annuli, and averaging the values within each shell to produce a single power spectrum measurement at the given length scale \citep{MoralesEOR1}.  This reduces the billions of individual power measurements down to of order ten statistical measurements, and is behind the extraordinary sensitivity of cosmological power spectrum measurements \citep{BowmanEOR3}.  

However, the first two stages in the foreground subtraction are not perfect.  For example, in the bright source removal stage the flux of each source will be slightly misestimated, leading to faint residual positive and negative sources at the locations of the subtracted sources. These faint residual sources inject spurious power to the three dimensional power spectrum.  It is impossible to determine what these subtraction errors are individually (otherwise we would improve them), however, we can predict, measure, and remove the \emph{average} effects of this residual foreground contamination from the power spectrum.  Since the power spectrum is related to the square of the intensity, residual positive and negative sources have the same power spectrum signature, and the amplitude of the residual power spectrum signal is related to the standard deviation of the subtraction errors made in the first two stages.  Different types of foreground subtraction errors produce distinct shapes in the three dimensional power spectrum, and are easily differentiated from the approximate spherical symmetry of the EOR signal ( see \citet{MoralesEOR1} and Section \ref{SubtractionErrorsSection}).  Figure \ref{symmetryFig} shows the three dimensional power spectrum shapes typical of the signal and residual foreground components.  

So in the presence of foregrounds our final stage of the analysis becomes a multi-parameter fit, with each component of the residual foreground and the EOR signal being represented by a corresponding 3D power spectrum template and amplitude.  The measurements are then decomposed into template amplitudes to separate the EOR signal from the residual contamination from foreground subtraction errors in the first two stages. In effect this final subtraction stage allows us to not only fit the local foreground parameters (position, spectra, etc.) as in the first two stages, but to also fit the width of the subtraction errors.

The errors produced by the first two foreground subtraction stages depend on the details of both the algorithm and the array.  For example, errors made in the spectral fitting stage depend on both the spectral model used (qudratic, power-law, etc.) and how the pixel shape varies with frequency (array design).  This precludes defining a set of residual error templates that is generally applicable, but calculating the templates for a specific analysis is straightforward, as demonstrated in the following section.

\subsection{Example Subtraction Error Templates}
\label{SubtractionErrorsSection}

To separate the residual foreground and cosmological signals in the residual error subtraction stage of the EOR analysis, we need to predict the shape of the residual foreground contamination as seen in the three dimensional power spectrum.  The first two stages of foreground subtraction remove the majority of the contamination, so what we see in the residual error subtraction stage is not the original power spectrum shape of the foregrounds, but instead the shape of the errors characteristic of the first two foreground removal stages. In the following subsections we provide examples of how to calculate the residual error templates, and discuss the characteristic power spectrum shapes.

\subsubsection{Statistical Spectral Fitting Errors}
\label{SpectralFittingErrorsSec}

In the spectral fitting foreground subtraction stage, a smooth spectral model is fit to each pixel to remove the contributions of faint continuum sources.  However, due to the presence of thermal noise the fit spectrum is not exactly the same as the true foreground. These slight misestimates of the foreground spectra in each pixel produce a characteristic power spectrum component. The exact shape of this power spectrum template of course depends on the spectral model one chooses.  Over the relatively modest bandwidths of proposed EOR measurements the foreground emission is reasonably well modeled by a quadratic spectrum \citep{BriggsForegroundSub}, and as an illustrative example we demonstrate how to calculate the power spectrum template in the case of a simple quadratic spectral model.

\begin{figure}
\begin{center}
\includegraphics[width=3.4in]{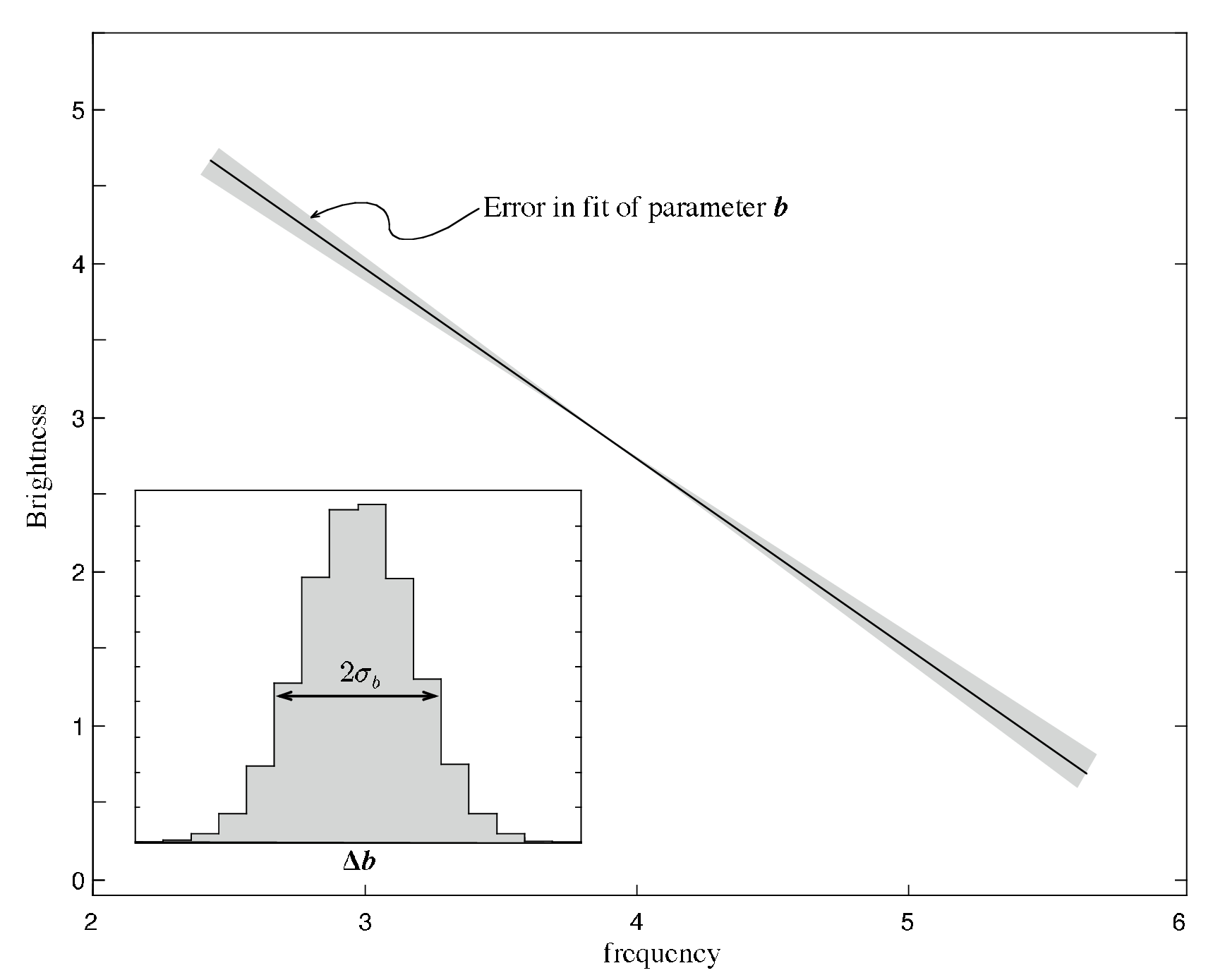}
\caption{This cartoon shows the true foreground continuum spectrum observed in one pixel as a black line, and the error in fitting parameter $b$ due to thermal noise.  The inset shows the expected Gaussian profile of $\Delta b \equiv (b_{T}-b)$ and the width of the distribution $\sigma_{b}$. }
\label{StatCartoon}
\end{center}
\end{figure}

For a quadratic foreground subtraction algorithm the residual foreground contamination is given by:
\begin{equation}
\label{ResidualEmissionEq}
\Delta S(f) = \Delta a\,df^{2 } + \Delta b\, df + \Delta c,
\end{equation}
where $df$ is the difference between the observed frequency and the center of the band, and $\Delta a, \Delta b, \Delta c$ represent the difference between the true parameter value for the foreground and the fit value.  Figure \ref{StatCartoon} depicts errors in fitting parameter $b$ for one pixel.
Moving to the line-of-sight wavenumber $\eta$ with a Finite Fourier Transform gives
\begin{equation}
\label{Seta}
\Delta S(\theta_{x},\theta_{y},\eta) =\frac{\Delta a B}{\pi \eta^{2}} - \frac{i \Delta b B}{\pi \eta} + \Delta c\delta^{k}(\eta),
\end{equation}
where $B$ is the bandwidth of the observation, and we have explicitly shown that this for a particular line of sight $\theta_{x},\theta_{y}$.\footnote{Equation \ref{Seta} assumes the function is sampled at the center of each bin, whereas in most measurements the value is integrated over the bin.  The two definitions converge as $\Delta a \rightarrow 0$, so we use Equation \ref{Seta} as a very good approximation.}

To compare with the three dimensional EOR power spectrum we need to move from the angular coordinates  $\theta_{x},\theta_{y}$ to the spatial frequencies $u,v$.  Even though the foreground is spatially clustered, we expect the fitting {\em errors} to be distributed like white noise.  This allows us to calculate the Fourier transform for the zero frequency term and generalize to the other values of $u$ and $v$:
\begin{equation}
\label{ }
\Delta S(u,v,\eta) = \sum^{\Theta}\Delta S(\theta_{x}, \theta_{y}, \eta) d\theta_{x} d\theta_{y}
\end{equation}
where $\Theta$ is the field of view and $d\theta_{x} d\theta_{y} = d\Omega$ is the angular resolution of the array (measured in steradians).  The root mean square of the sum is given by the square root of the number of independent  lines of sight ($\sqrt{\Theta/d\Omega}$) times $d\Omega$
\begin{equation}
\label{ }
\Delta S(u,v,\eta)_{rms} = \Delta S(\theta_{x}, \theta_{y}, \eta)_{rms} \sqrt{\Theta d\Omega}.
\end{equation}

If the errors in the fitting parameters are Gaussian distributed, the average power spectrum will be 
\begin{equation}
\label{ResidualSpectrumPSEq}
\langle P_{s}(\eta) \rangle = 2 \Theta d\Omega B^{2}\left[ \frac{\sigma^{2}_{a} }{\pi^{2} \eta^{4}}  + \frac{\sigma^{2}_{b}}{\pi^{2}\eta^{2}} + \sigma^{2}_{c'}\delta^{k}(\eta)\right],
\end{equation}
where $\sigma_{a}$ and $\sigma_{b}$ are the standard deviations of $\Delta a$ and $\Delta b$ respectively, and the term $\sigma_{c'}$ has been re-defined to include all the contributions proportional to the Kronecker delta function $\delta^{k}(\eta)$.   For each visibility $u,v$, the power spectrum will be exponentially distributed around the average, but will become Gaussian distributed when many lines-of-sight are averaged together.

\begin{figure}
\begin{center}
\includegraphics[width=3.4in]{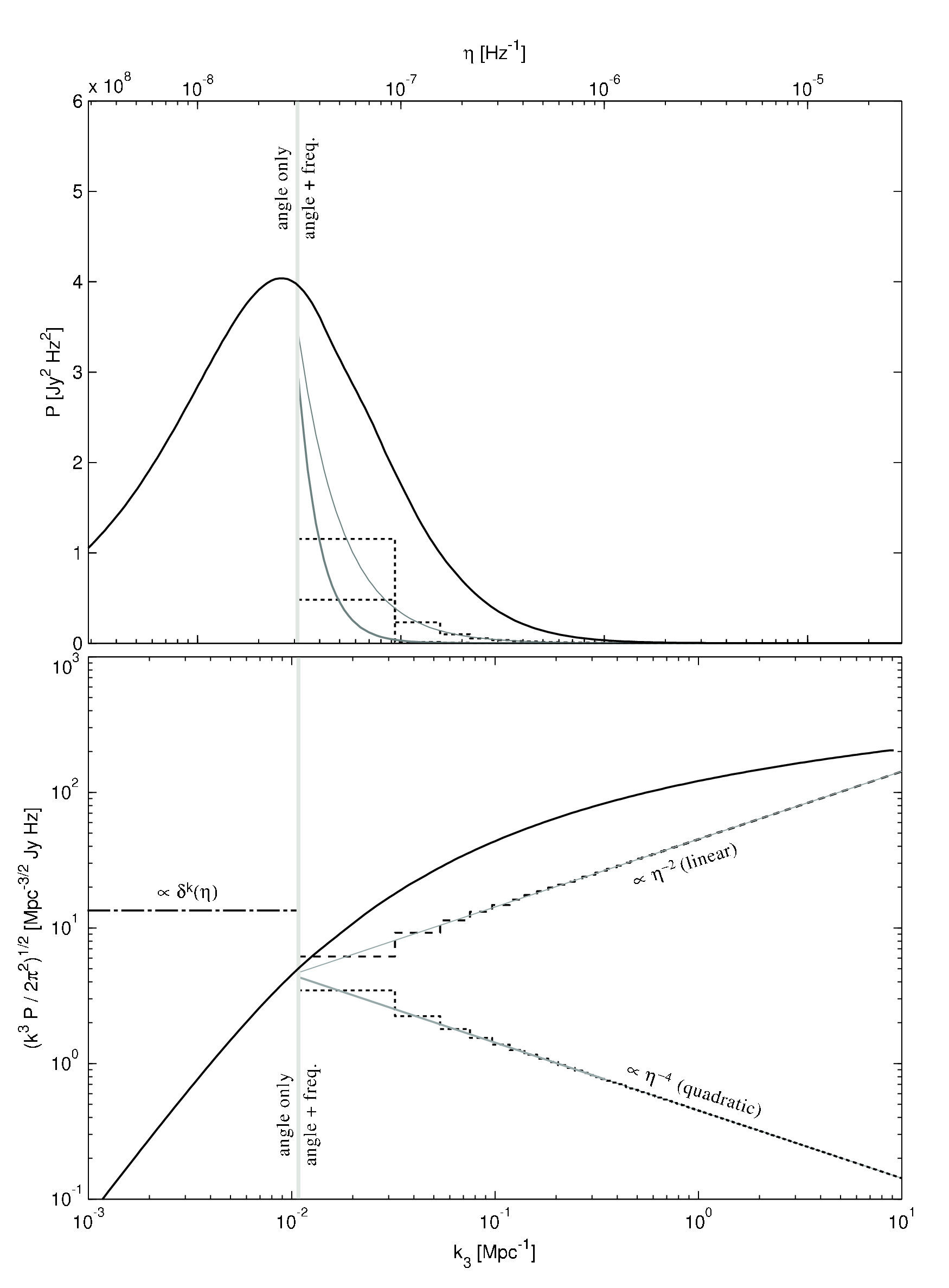}
\caption{This figure shows the shape of the power spectrum contributions from the EOR signal and the statistical spectral fit residuals in observer's units (top panel) and theorist's units (bottom panel) along the line-of-sight direction ($\eta$ or $k_{3}$).  The observed EOR signal is shown as a solid black line, and the linear ($\propto \eta^{-2}$) and quadratic ($\propto \eta^{-4}$) residuals are shown as thick and thin grey lines respectively. The vertical grey line shows the edge of the $\eta = 0$ bin for a 16 MHz bandwidth.  All power spectrum measurements to the left of the vertical line are purely angular while the measurements to the right of the line use both the angular and frequency information. The dashed lines show the binning effects for the linear and quadratic components of the residual foreground, with the dash-dot line showing the $\delta^{k}$-function contribution from the offset term.  The amplitudes of the residual foreground components depend only on the standard deviations of the fitting parameter errors ($\sigma_{a}, \sigma_{b}, \sigma_{c'}$) and are fit using parameter estimation in the residual subtraction stage of the analysis. }
\label{StatFourier}
\end{center}
\end{figure}

Equation \ref{ResidualSpectrumPSEq} gives the average power spectrum due to statistical fitting errors with a simple quadratic spectral model. A plot of this fitting error power spectrum template along the line-of-sight is shown in Figure \ref{StatFourier}.  For pixel-based fitting algorithms the thermal noise is nearly constant for all pixels, and thus the magnitude of the residual power spectrum is nearly equal for all visibilities.  This provides a particularly simple three dimensional power spectrum template that falls as a high power of $\eta$ and is constant for all visibilities, and can be visualized as a plane of power near $\eta = 0$ that quickly falls away (see the right hand panel of Figure \ref{symmetryFig} for the basic shape). For visibility-based subtraction algorithms tuned for larger scale structure \citep{ZaldarriagaPow1}, the thermal noise at large angular scales is much lower due to the central condensation of realistic arrays,  so the power will be concentrated near the $\eta = 0$ plane but reduced in amplitude at small visibilities.

The $\sigma$ values in Equation \ref{ResidualSpectrumPSEq} represent the standard deviation of the quadratic spectral fitting algorithm and determine the amplitude of the residual subtraction errors to the signal.  The $\sigma$ represent an ensemble statistic, as compared to the individual fits to each pixel made by the spectral fitting algorithm.  In the full analysis we fit local parameters for each pixel/visibility during the spectral fitting stage, and then we fit the errors ($\sigma$) in the residual foreground subtraction stage.  This pattern of fitting local parameters in the first two stages, and the statistical distribution of their errors in the third is repeated for each type of error.  

The example power spectrum template presented here only applies to a simple quadratic spectral model, but similar results can be obtained if power-law or other foreground models are used instead.  They key is to determine the shape of the power spectrum produced by the local spectral fitting errors, which can then be fit globally in the residual error subtraction stage of the analysis.

\subsubsection{Spectral Model Errors}

\begin{figure}
\begin{center}
\includegraphics[width=3.4in]{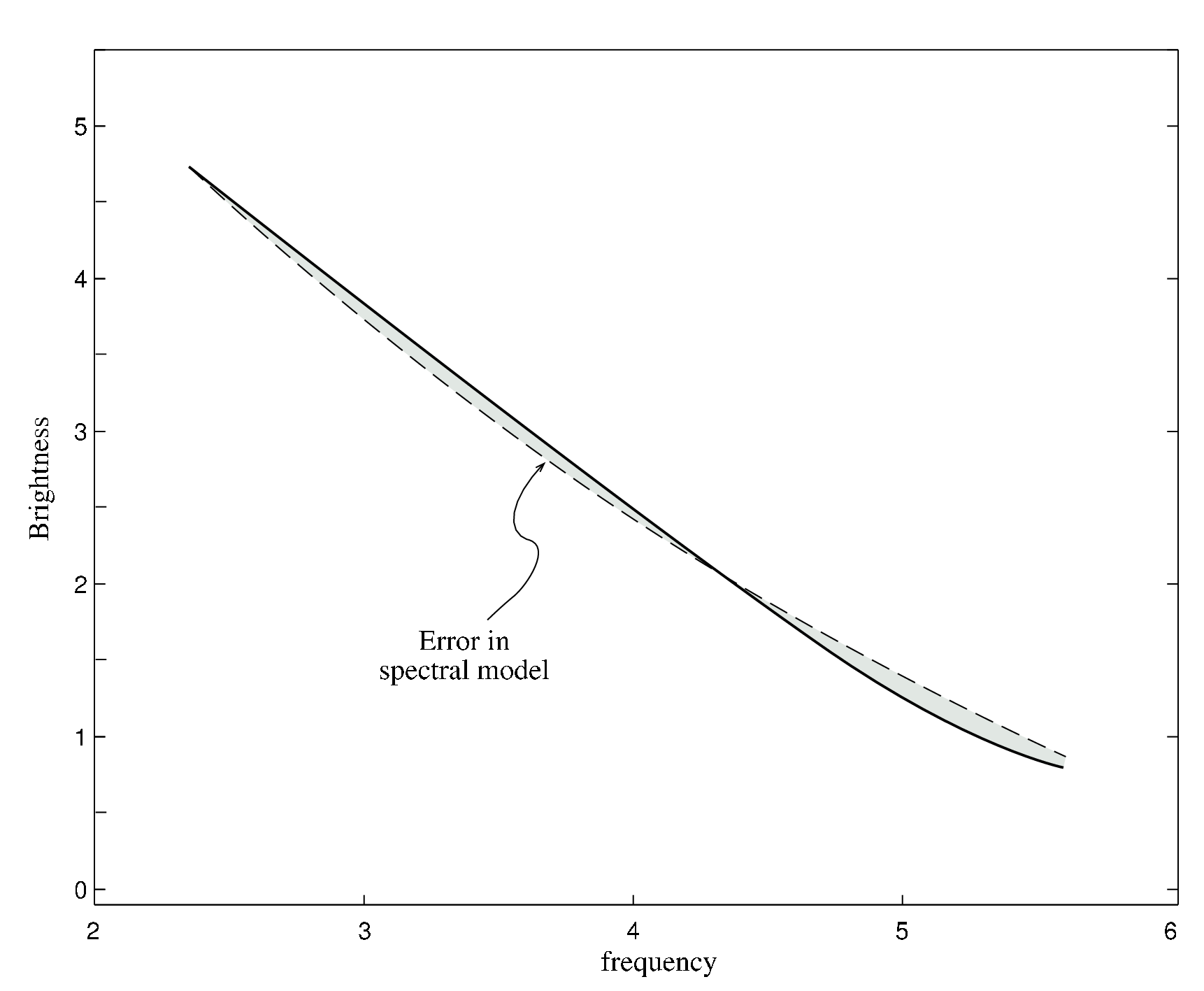}
\caption{This cartoon shows the difference between the spectrum in a pixel (solid line) and the best fit model in the absence of measurement noise (dashed line).  Because the model cannot exactly follow the true spectrum, an additional component is added to the power spectrum which can be estimated and removed in the residual error subtraction stage of the analysis.}
\label{ModelError}
\end{center}
\end{figure}

In addition to the statistical fitting errors discussed in the previous subsection, there is a class of model errors which can be made in the spectral fitting stage.   Simple foreground spectral models may be unable to fit the underlying foreground spectrum.  Even if all the source spectra were perfect power laws, there are many sources per pixel, which can lead to a complex cumulative spectrum which cannot be exactly fit by a simple spectral model.  Figure \ref{ModelError} shows the origin of the model errors. 
Because the summation of sources and spectra that forms the cumulative spectrum in each pixel is a statistical process, we expect the model errors to be statistical.  This allows one to follow the same process we used in the previous section and form an expected power spectrum shape due to model errors which can be fit in the residual error subtraction stage of the analysis. This model error template will depend sensitively on both the chosen spectral model and the imaging characteristics of the array.

\citet{BriggsForegroundSub} showed that using realistic brightness counts $dN/dS$ and spectral slopes, the cumulative spectrum is typically dominated by the brightest source in the pixel, and even for a simple quadratic spectral fit the model error is expected to be less than the cosmological EOR signal.  Thus we expect the power added by model errors to be quite small.  But by including the model error power spectrum in the parameter fit, we can eliminate any bias that could be introduced into the EOR signal.

\subsubsection{Bright Source Subtraction Errors}
\begin{figure*}
\begin{center}
\plotone{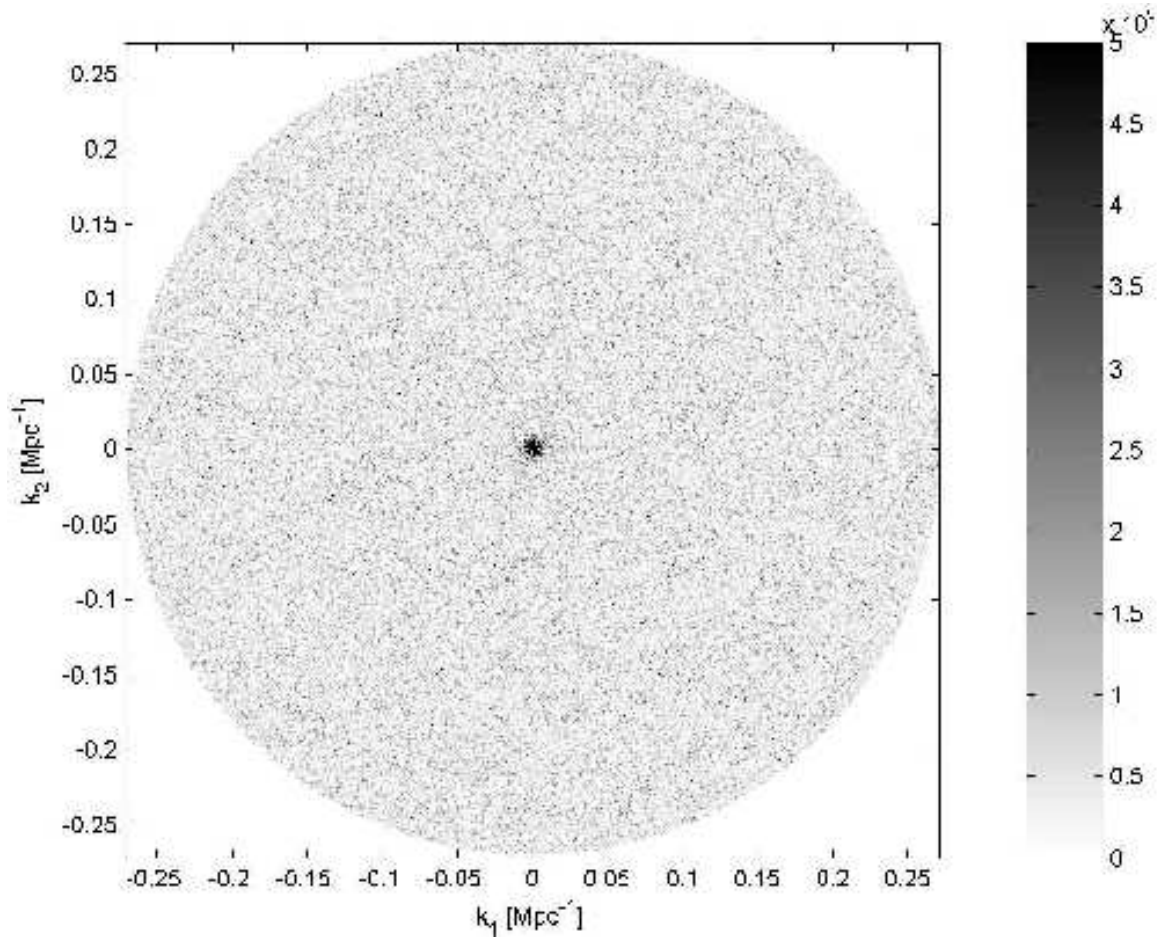}
\caption{This figure is an example of the expected $u,v$ power spectrum template due to bright source subtraction errors.  In this example 14,510 sources from the Westerbork Northern Sky Survey (WENSS) at 325 MHz \citep{WENSS} were chosen in a field centered at 90 deg RA, 60 deg DEC, with a 31 degree FOV(diameter).  To model the compression of the dynamic range, the variance of the source subtraction error was chosen to be proportional to the log of the flux, and the color map is linear with an arbitrary scale.  Note the increased amplitude towards the center due to the angular galaxy correlation. The exact power spectrum template will depend on the chosen field and the source identification algorithm, but will have this basic shape in the $u,v$ plane and will be highly concentrated towards $\eta = 0$.}
\label{brightErrorFig}
\end{center}
\end{figure*}
Astrophysical sources which are bright enough to contaminate distant areas of the image cube are removed in the bright source subtraction stage of the analysis (Section \ref{BrightSourceSec}).  The errors in this foreground removal stage are primarily due to misestimates of the source intensities.  The bright source subtraction errors can be envisioned as residual positive and negative sources at the locations of the subtracted sources, and will produce a distinct foreground signature in the three dimensional power spectrum.

The accuracy of determining the fluxes of foreground sources depends on their brightness---the strongest sources may be subtracted to a few parts per billion while faint sources are only subtracted to a few tenths of a percent accuracy.  Thus the residual sources mirror the positions of subtracted foreground sources and have a zero mean Gaussian distribution of flux, but with a greatly reduced dynamic range (the standard deviation of the subtraction error $\sigma$ is \emph{not} proportional to the brightness $S$).  This produces a map of the expected residual sources which can be convolved with the point spread function of the array and Fourier transformed and squared to create the expected three dimensional power spectrum template for the errors introduced by the bright source subtraction stage of the analysis.

The source subtraction error template will depend on the particular field observed and spectral model (see Section \ref{SpectralFitSec}), but will be strongly peaked at low $\eta$ (primary error leads to $\delta^{k}(\eta)$) and concentrated at small visibilities due to angular clustering of the foreground sources \citep{DiMatteoForegrounds}.  The compression of the dynamic range must be either modeled, or fit with an additional parameter in the residual error subtraction stage of the analysis. An example power spectrum template is shown in Figure \ref{brightErrorFig} using sources from the Westerbork Northern Sky Survey (WENSS).  As can be seen from the example, the errors made by the bright source subtraction stage add power, particularly at small $u,v$ and $\eta = 0$.  A similar procedure can be used for RFI which spills into neighboring frequency channels, and depends on the spectral dynamic range of the array and the RFI removal procedure.

\subsection{Uncertainty Calculations}
\label{UncertaintySec}

In the final residual foreground subtraction stage, parameter estimation is used to separate the EOR signal from the residual foreground contaminants, using the characteristic power spectrum templates calculated in Section \ref{SubtractionErrorsSection}.  We are left with a measurement of the EOR signal strength (typically in ranges of $k$) and the residual foreground amplitudes.  The uncertainty of the EOR signal determination depends on two factors:  the uncertainty in the residual foreground amplitude determinations, and their covariance with the EOR signal.

Calculating the uncertainty of the foreground amplitudes ($A_{x} \propto \sigma_{x}^{2}$) is complicated by their additional statistical correlations as compared to the EOR signal. The observed Fourier space has a fundamental correlation imprinted by the field-of-view and bandwidth of the observation. This can be used to define a natural cell-size for the data where the observed EOR signal in each cell is largely independent \citep{MoralesEOR1,MoralesEOR2}.  The observed data can then be represented as a vector of length $N_{u}\times N_{v} \times N_{\eta}$, where $N$ is the number of cells in the $u,v$ and $\eta$ dimensions respectively.  For the EOR signal vector $\mathbf{s}$, all of the cells are nearly independent and $\langle\mathbf{ss}\rangle \approx \mathbf{sI}$ where $\mathbf{I}$ is the identity matrix.  

This is \emph{not} true for most of the residual foreground templates.  The residual foreground contributions are often highly correlated between Fourier cells ($\langle\mathbf{ff}\rangle \not\approx \mathbf{fI}$), and so average differently than the EOR signal.  For example, the linear and quadratic spectral fitting errors ($\sigma_{a}$ and $\sigma_{b}$ terms) each imprint a specific residual in all the $\eta$ channels for a given $u,v$ pixel --- the amplitude of the residual will vary from one $u,v$ pixel to the next, but are perfectly correlated (deterministic) for the $\eta$ values within one pixel. 

When averaging over many Fourier cells and lines-of-sight, the uncertainty in the amplitude of any component becomes approximately Gaussian distributed and equal to
\begin{equation}
\label{amplitudeUncertainty}
 \Delta A_{x} \equiv A_{x}^{T} - A_{x} \propto \sqrt{\frac{A_{x}}{N_{m}}},
\end{equation}
where $A_{x}$ is the amplitude of the $x$ component of the signal or residual foreground contribution, $^{T}$ is the true value, and $N_{m}$ is the number of independent measurements (realizations) of this contribution.  For the example of the $\sigma_{a}$ and $\sigma_{b}$ terms of the spectral fitting residual errors, there are only $N_{m} = N_{u}\times N_{v}$ independent measurements as compared to $N_{u}\times N_{v} \times N_{\eta}$ for the EOR signal and the thermal noise.  This correlation along the $\eta$ axis comes from using the frequency channels to make the original spectral fit, and means that these spectral errors will only average down by adding more lines of sight, not increasing the number of cells along the $\eta$ axis. The correlations of the various foregrounds are listed in Table \ref{CorrelationTable}.

\begin{deluxetable*}{cccc}
  \tablehead{& Functional Dependence & Statistical Correlations & $N_{m}$ }
  \startdata 
  EOR Signal & $\mathbf{k}$ & --- & $N_{u} \times N_{v} \times N_{\eta}$\\
  
  Thermal Noise & $\sqrt{u^{2}+v^{2}}$ & --- & $N_{u}\times N_{v} \times N_{\eta}$ \\
  
  Spectral Fitting ($\sigma_{a},\sigma_{b}$) & $\eta$ & $\eta$ & $N_{u}\times N_{v}$ \\
  
  Spectral Fitting ($\sigma_{c'}$) & $\delta^{k}(\eta)$ &  --- & $N_{u}\times N_{v}$ \\
  
  Spectral Model & $\eta$ & $\eta$ & $N_{u}\times N_{v}$\\
  
  Source Subtraction & $u,v,\eta$ & $u,v,\eta$ & $N_{\mathrm{sources}} \times N_{\mathrm{spectral\ parameters}}$
  \enddata
  \tablecomments{The table shows the functional dependence and statistical correlations of the signal and foreground components in the Fourier space ($\mathbf{k}$ or $u,v,\eta$), and the number of independent measurements of the component's amplitude $N_{m}$. The observed Fourier space has a fundamental correlation imprinted by the field-of-view and bandwidth of the observation, and this can be used to define a natural cell-size for the data set where the observed EOR signal and thermal noise in each cell is independent \citep{MoralesEOR1,MoralesEOR2}.  However, the residual foreground components have additional statistical correlations between cells, as indicated in the table and described in the text. }
  \label{CorrelationTable}
\end{deluxetable*}
In addition to the uncertainties in the amplitude determinations, we must also calculate the covariance of the parameters.  Since the power spectrum templates of the residual fitting errors and signal form the basis functions for the parameter estimation in the residual error subtraction stage, they define the covariance of the amplitude terms. Thus, for a given observatory and choice of foreground subtraction algorithms, we can calculate the residual foreground power spectrum templates and resulting uncertainty in the EOR measurement.

\section{Implications for Array and Algorithm Design}
\label{ConstraintsSection}

Since the power spectrum templates of the subtraction errors quantify the interactions between the three analysis stages, they enable us to study the effects of array design on our ability to isolate the statistical EOR signal.  The difference in the shapes of the power spectrum templates determines how easy it is for the parameter estimation stage of the analysis to separate different contributions.  If the power spectrum shapes of two contributions are similar, it will be difficult to accurately determine the amplitudes of the contributions. Mathematically, the shapes of the power spectrum templates determine the covariance of the parameter estimation matrix, with the covariance decreasing as the shapes become more orthogonal.  

The power spectrum templates depend on the details of both the array design and analysis technique.  For example, the model fitting error template depends on both the angular resolution of the array (detailed pixel shape) and whether a quadratic or logarithmic power law is used in the fit.  The performance advantages and trade-offs of different arrays and analysis techniques is captured in the shapes and covariances of the power spectrum templates. 

To date, the experimental community has been uncertain as how to best design arrays and analysis systems to detect the statistical EOR signal.  Much of this confusion is because no quantitative measure has been available for comparing design choices.  We feel that the power spectrum templates can provide the necessary figure of merit.  The shape of the EOR power spectrum signal is known (given a theoretical model), and so we are concerned with the amplitude and covariance of the foreground subtraction errors of a given array and analysis system with the known EOR power spectrum.  The power spectrum templates define the performance of the array and analysis, and allows design trade-offs to be accurately compared.  In many cases, making a plot analogous to Figure \ref{StatFourier} and comparing the amplitude and shape of the residual templates will be sufficient to guide the array design.

\section{Towards Precision Foreground Calculations}
\label{ForegroundModels}

The residual foreground contamination levels shown in Figure \ref{StatFourier} are unrealistically small for the first generation of EOR observatories.  However, the EOR signal can still be detected even if the amplitude of the residual foregrounds greatly exceeds the EOR signal in the Figure.  The key question is not the amplitudes of the residual foregrounds, but the uncertainties they create in measuring the EOR signal, as discussed in Section \ref{UncertaintySec}.  The foregrounds shown in Figure \ref{StatFourier} are what would be needed to detect the EOR power spectrum in a single pixel, whereas all of the first generation EOR observatories rely on combining information from many lines of sight.

Unfortunately, the uncertainty due to foreground contamination depends strongly on characteristics of the array and observing strategy, for example, a wide field observation will average over more lines of sight and thus be able to tolerate higher standard deviation in the subtraction than a narrow field observation. This precludes defining a set of amplitudes which must be obtained to observe the EOR signal for a generic observatory.

The dependence of the subtraction precision on the details of the array makes the task of foreground modelers much more difficult.  The precision of the foreground removal is now array dependent, and most researchers are not familiar with the subtle array details needed to accurately calculate the sensitivity of a given observation.

However, the power spectrum templates and $\sigma$ values do offer a way of translating from the characteristics of a foreground removal algorithm to the sensitivity of an array.  Modelers can determine the shape of the residual foreground contamination (as in Equation \ref{ResidualSpectrumPSEq}) and the scaling of the $\sigma$ values for their foreground removal algorithm.  Experimentalists can then use the predicted power spectrum shapes and $\sigma$ scalings to determine the effects of the foreground subtraction for a specific observation.  This allows researchers studying the foreground removal to avoid doing detailed calculations for each array and observing strategy, while still providing robust results which can guide the experimental design of the next generation EOR observations. In this way, we hope the foreground removal framework presented in this paper will facilitate a conversation between foreground modelers and experimentalists and enable accurate array-specific predictions of the foreground subtraction effects on the up coming EOR observations.

 \end{document}